\documentclass[11pt]{article}
\pdfoutput=1
\usepackage{amsmath}
\usepackage{amsfonts}
\usepackage{latexsym}
\usepackage{graphicx}
\usepackage{color}
\usepackage[colorlinks,citecolor=magenta,bookmarksopen,bookmarksdepth=2,breaklinks=true]{hyperref}
\usepackage{fullpage}
\usepackage[normalem,normalbf]{ulem}
\usepackage{natbib}
\usepackage[section]{placeins}
\usepackage{times}


\usepackage[markup=bfit]{changes}
\definechangesauthor[name={Subhashis Banerjee}, color=orange]{suban}
\setremarkmarkup{(#2)}
\usepackage[shadow]{todonotes} 

\title{Authorisation and access control architecture as a framework for data and privacy protection}

\author{\href{http://www.cse.iitd.ac.in/~suban}{Subhashis Banerjee} \\ Computer Science and Engineering \\ IIT Delhi}

\date{30 January, 2018; updated on \today} 

\begin{document}
\maketitle

\begin{abstract}
Privacy protection in digital databases does not demand that data should not be collected, stored or used, but that there should be guarantees that the data can only be used for pre-approved and legitimate purposes. We argue that a data protection law based on traditional understanding of privacy protection and detection of privacy infringements is unlikely to be successful, and that what is required is a law based on an understanding  of  the architectural requirements of authorisation, audit and access control in real-time.  Despite the protection principles being sound, privacy protection in digital databases has been less than effective, anywhere,  mainly because of weak enforcement methods.
\end{abstract} 

\section{Introduction}
The debate engendered by the identity project has suddenly propelled us from being a predominantly pre-privacy society to one in which privacy protection in digital databases has emerged as a major national concern. The welcome and scholarly Supreme Court judgment on the right to privacy \citep{puttaswamy, bhatiaepw}  has made it abundantly clear that privacy protection is imperative, and any fatalistic post-privacy world view \citep{post-privacy} is untenable. Informational self-determination and the autonomy of an individual in controlling usage of personal data have emerged as  central themes across the  privacy judgment. See \citep{Bhatia4} for a summary.

Recording transactions with a digital identity projects an individual into a data space, and any subsequent loss of privacy can happen only through the  data pathway. Hence data protection is central to privacy protection insofar as digital databases are concerned. Two main uses of digitisation of transactions are accurate record keeping for audit and post facto investigation, and data analytics to discover useful patterns in the data. The latter can be particularly useful for econometric analysis, epidemiological studies and latent topic discovery. It can hence facilitate improved design of social policy strategies and early detection and warning systems for anomalies \citep{SSS}. However, both record keeping and analytics can lead to potential loss of privacy. Long term storage of digital records, indexed by an identity, may result in unauthorised accesses which may violate personal data autonomy. Analytics often require joining of data from different domains to be able to mine useful knowledge. Such breaking of silos, in turn, can also be used to profile  individuals  beyond legal sanctions and can thus violate the principle of informational self-determination. Hence, effective use of digital databases and privacy protection have some apparently contradictory requirements. 

In this note we investigate the possible ways in which such contradictions may be resolved. We also analyse the recommendations of the Srikrishna committee on a data protection framework \citep{BNSWP}. We argue that any solution that is solely based on detection of privacy violation and subsequent remedial action is unlikely to be satisfactory, and what is required in addition is an online architectural solution that  prevents privacy invasions in the first place. As pointed out in \citep{malavika},  {\em ex-ante} rather than {\em ex-post} should be the preferred approach.

The welcome emphasis on data privacy provides us with a unique opportunity to take a fresh look at protocol design for effective digital services in India. On the one hand our systems should offer stricter privacy protection than what is prevalent in the US, where not only are identity theft rates unacceptably high \citep{LSE}, but also where some of the world's largest corporate panopticons like Google and Facebook have grown more or less unchecked. On the other hand  India should ideally have a more innovation friendly setup than what the  European GDPR \citep{gdpr} can offer, which perhaps is unduly restrictive but is unlikely to be commensurately effective. Moreover, our designs  need to be specially sensitive to our large under-privileged population which may not have the necessary cultural capital to deal with  overly complex  digital setups.

\section{Privacy risks in digital databases}

The first step to privacy protection is to understand the ways by which privacy may be compromised in digital databases.

The most common fear of  digitisation, especially when enforced by governments, is that of mass surveillance. All digitisation require unique digital identities, and the use of a universal identity in a multitude of databases can possibly create an infrastructure for totalitarian observation of citizen's activities across different domains. It may be naive to repose faith in benevolence of governments, anywhere, and the mere presence of an unrestricted surveillance infrastructure, which digital databases linked by  unique identities can easily lead to,  can potentially disturb the balance of power between the citizens and the state, stifle dissent and  be a threat to civil liberty and democracy \citep{dreze}. Several commentators have used  cliched metaphors like {\it Orwellian big brother} or {\it panopticon} to describe the situation.

A far more common and subtle manner of erosion of privacy is by the way of losing control of informational self-determination both to the state and to other seemingly mysterious,  uncaring and opaque bureaucracies. Often there is no direct or obvious invasion of privacy, but because of the ease of replication, aggregation and selective combination of whole or parts  of personal digital data,  one may sometimes become unsure about what information about her is being used by the state and other bureaucracies and for what purposes. \citep{solove} argues that  {\it Kafkaesque} is a more appropriate metaphor for describing this situation. Not only can personal information leach out and be used in unpredictable ways by unpredictable entities, but one can also  be mis-profiled, wrongly assessed or even influenced \citep{cambridgean} using out of context data, without being able to control such decisions or sometimes even being aware of them.

Yet other typical pitfalls of bureaucratic digitisations are poorly thought out use cases, incomplete case analyses and incompetent programming. As common fallouts one may suddenly find herself being  deregistered from services due to no fault of her's, or having to unnecessarily run around to get things corrected when she was not the one responsible for the mistakes in the first place.   Being denied participation in school painting or dance competitions because of want of a national digital identity, or being denied hospital treatment, pension or welfare because perhaps a name is misspelt or because fingerprints do not match will be cases in point. Such callous omissions can not only be threats to right to privacy, but, in extreme cases,  even to rights to liberty and life \citep{santoshi}. 

Finally, threats to privacy and liberty  also arise from big-data analytics or machine learning algorithms, which are important reasons for collecting and recording high frequency, real-time and non-aggregated transactional data in the first place.   \citep{Cathy} forcefully argues that big-data analytics, by the very fact that they are designed  by the privileged and often for profit,  ``increases inequality and threatens democracy'' (see \citep{Cathyreview} for a review). She illustrates with a series of examples ranging from  assessment and estimation of teacher quality, recidivism risk, creditworthiness and college rankings to employment application screeners, policing and sentencing algorithms and workplace wellness programs to show that they reinforce inequality and reward the rich and punish the poor. The bias is either present in the algorithm or in the data and sometimes even in both. The common traits of such poor fallout of predictive analytics usually are opacity, scale, and damage. \citep{sagepub} argue that ``ÒAs algorithms select, link, and analyse ever larger sets of data, they seek to transform previously private, unquantified moments of everyday life into sources of profit.''

Despite the above risks, because of the enormous benefits that digitisation and analytics promise,  they are faits accompli and the question is only one of realising them safely and well. The {\it Orwellian big brother} and the {\it Kafkaesque} arguments certainly raise crucial concerns, but they do not necessarily imply that privacy protection is impossible with digitisation using a unique identifier. Also, despite clearly suggesting that ``The technology exists! If we develop the will, we can use big-data to advance equality and justice'' \citep{Cathy},  her work is often interpreted to infer that predictive algorithms are necessarily evil. Such a conclusion, from a few examples of badly done predictive analytics, is  overly pessimistic and inductivism at its worst.

\section{Analysis of suggested measures}

Attempts at privacy protection in digital databases have mainly been based on the tenets of legitimate state interest; informed consent and notice; collection, purpose and storage limitation; participation of individuals; transparency; regulations, enforcement and accountability \citep{shah,BNSWP}. These measures, however, have turned out to be less than effective in preventing a shift to a post-privacy world, at least insofar as personal data is concerned. We try to understand why?

\subsection{Legitimate state interest}
Clearly, the same privacy protection principles cannot be horizontally applied to the state and other essential bureaucracies, like banking and insurance  for example, and to non-essential private digital services where user participation is voluntary. In the first situation the state would most often require the digitisation to enforce {\em compliance}, such as in income tax. In all such cases the state can mandate digitisation of personal information only after establishing a {\it legitimate state interest} and enacting a law \citep{puttaswamy,barandbench}. All constitutional tests of proportionality, reasonableness  and non-arbitrariness would need to be applied. Here, the role of  {\em consent}  would be minimal, but {\em collection} and {\em purpose limitation} would be important operative principles. However, merely enacting a law would not absolve the state and other bureaucracies from the responsibility of protecting  privacy rights of individuals, and all the risks mentioned in the previous section will  still have to  be mitigated. The state's understanding of this principle, however, is often questionable.

\subsection{Informed consent, notice, purpose limitation and opt-out}
For all other cases of voluntary participation, {\em privacy self-management} \citep{solove2} through {\em informed consent} operationalised by effective  {\em notice};  {\em collection, purpose and storage limitation}; {\em transparency}; and {\em individual participation} through {\em opt-in} and {\em opt-out}  have often been advocated as  foundational principles for privacy protection \citep{shah}. However, as pointed out in \citep{BNSWP,solove,matthan}, 
 notice and consent are usually ineffective because of  information overload, limited choice and {\em consent fatigue}. In fact, the customary negligent clicking of `I Agree' and the overwhelming burden of information required for informed consent, both in terms of volume and complexity, have made the consent principle impractical for privacy protection. \citep{matthan} makes a strong case for a rights-based approach that shifts a significant part of the responsibility and accountability  from the individual to the data controller, irrespective of the level of consent. This  clearly is required, in addition to consent, and a strong regulatory framework can lay out the standards. However, the methods of enforcement and detection of breaches remain open questions.
 
Similarly, it may not always be possible to enumerate the purposes for which personal data may be used at the time of collection, and {\em purpose limitation} is a difficult privacy protection principle to administer \citep{pl1}. In particular, any inflexible implementation of purpose or storage limitation can severely impede innovation in the age of predictive analytics and machine learning where new uses of transactional data and new methods of processing are being discovered every day. Perhaps a combination of {\em legitimate interest} and purpose limitation under reasonable regulatory control is what is required. However, the regulatory control should not be so lax that it is ineffective, neither should it be so overbearing or paralytic with inertia that it stifles innovation.

Finally, whenever there is a purpose extension not covered by a legitimate state interest, there should always be a notice for consent renewal and an  opt-out alternative with a guarantee of deletion of all personal data. It  is the recognition and acknowledgement of purpose extensions, however, that have often been problematic. 

\subsection{Right to explanation}

The European GDPR proposes right to explanation as a countermeasure to indiscriminate and biased machine learning applications \citep{ipx022}. However, predictive analytics rarely support causal reasoning, and, without expert audit of algorithmic and data biases, the explanations will most likely turn out to be inane. Moreover, the adverse outcomes of perverse machine learning applications are {\em Kafkaesque}, and the consequent damages are not immediately obvious. So timely explanations may never even be sought.

\subsection{Regulation and enforcement}
\citep{BNSWP} propose a strong regulatory framework for enforcement of privacy standards and for fixing accountability. The framework can range from the currently prevailing  self-regulation to a more prescriptive ``command and control", and the committee advocates a middle path of co-regulation. However, the approach presupposes that privacy invasions are detectable. The `data as property' view for privacy protection espoused by \citep{Tarafder} also assumes the same. This assumption, however, is problematic since detection of  privacy infringements, especially of the {\em Kafkaesque} types, will always be uncertain because the causal effects of invasions  will be hard to determine. For example, it may turn out to be impossible to know for sure whether a person has lost her job because her personal medical data was accessed without authorisation and  used to discriminate against her, or some other reason put out as the official explanation was really the determining factor. 

We propose instead that the regulatory framework be built into the privacy protection architecture \citep{SSS}.

\section{Elements of an architectural solution}

Pivotal to protecting the autonomy of an individual is to protect her identity from getting disclosed when not required,  to protect her from unauthorised profiling without her knowledge, and to inform her about all accesses to her data and their outcomes. We contend that the following architectural properties in digital databases are crucial to achieve these objectives.  

\subsection{Identity protection through virtual ids}
Using the same personal identifier for all applications and frontend databases is architecturally unsound from a privacy protection point of view.  A straightforward alternative would be to use different virtual identifiers for each application domain, making unauthorised correlation of identities across silos impossible. The mapping between the different virtual identifiers can be maintained securely at a central place with strong access control protocols, and can be used to facilitate legal and authorised mining of personal information across silos \citep{SSS}. This would be required not only for a national digital identity, but also for all other unique personal identifiers like the income tax PAN or  mobile phone numbers, which have tacitly been converted to unique digital identifiers by the Indian private enterprise. There should be no need to disclose one's real phone number to  vendors or for  train journeys, and it should be possible to generate virtual and limited duration numbers linked to the original ones on demand. A backend mapping can then route calls and messages to the real ones.

 In addition, for effective privacy protection it will be imperative not to use any other weak identifiers like names and addresses in frontend databases. All frontend transactions should store only virtual numeric identifiers after verifying the authenticity using a safe protocol.  It is well known that anonymisation with  provable guarantees \citep{difforivacy} is hard to achieve in presence of weak identifiers.

\subsection{Online regulatory framework for authorisation and access control}

Not only are independent regulatory authorities overseeing the data controllers necessary for privacy protection, we argue that the regulatory authorities need to have active presence in the data protection architecture to enable them to prevent privacy breaches from happening. Apart from grievance redress and determining fairness of algorithms and use cases, they need to play two other main roles.

The first role ought to be to determine and clearly define who can access what data and for what purposes, based either on legal sanction or on a consent principle, in conformance with a rights-based data protection law.  Purpose limitation needs to be built into such authorisations, and all purpose extensions and consent renewals should be explicitly considered. Such access rights to personal data should not only be defined for operational, investigation and audit purposes, but also for granting algorithmic access for data mining. As \citep{sethepw} points out, despite the recent  progress in attempts to build fairness into algorithm design, fairness guarantees are not always possible \citep{kleinberg}. Hence manual scrutiny and regulatory control of both  algorithmic procedures and their use in societal applications are imperative.  

Once such access rights are clearly defined and digitally coded, and the authorisations recorded, the second crucial  role should be to ensure that data can be accessed only through audited, pre-approved and digitally signed computer programs after online authentication and verification of the authorisations presented.
 The regulators should also ensure that the accessed data is used only for  authorised purposes by authenticating the genuineness of the accessing programs in real-time. This would require the regulators to audit, approve and digitally sign all programs that  data controllers may use to access and process the data. Both the data regulator and the data controller should maintain non-repudiable logs of all data accesses, and neither should be able to access the data independent of the other. 
 Finally, for the sake of transparency, all outcomes of accesses to personal data should always automatically be communicated to the concerned individuals through private channels.
 
The technology to support such regulatory functions exists (see \citep{SSS} for a preliminary analysis), what are necessary now are the will to build the required regulatory capacity and an effective, rights-based data protection law. A co-regulatory approach, where private bureaucracies  and data controllers can have their own independent data regulators who can act like online ombudsmen and monitor and enforce privacy protection, can possibly help in building such regulatory capacity faster. Essential state bureaucracies like the national identity or income tax authorities will however require a central data regulatory authority.

\section{Conclusions}
We have argued that a passive regulatory framework based on detection of privacy breaches, and traditional approach to privacy protection based on the principles of consent, purpose limitation and transparency is unlikely to be successful. In addition to these standard measures we advocate an architectural solution based on online validation of authorisation and access control to prevent privacy infringements in the first place.

\section*{Acknowledgement}
I thank all students and  faculty colleagues who participated in the course on {\em Digital infrastructure, identity, online data and privacy} held between September and November, 2017 at IIT Delhi.  I also thank the guest speakers Usha Ramanathan, Elizabeth Bennett and Arghya Sengupta for their insightful talks in the course. I specially thank Subodh Sharma for his many useful comments on this manuscript.
\bibliographystyle{plainnat}

\bibliography{dp}


\end{document}